\documentclass{article}
\usepackage[utf8]{inputenc}
\usepackage{graphicx, amsmath,amssymb,algorithm, 
algpseudocode}
\usepackage{amsthm, alltt, amssymb, xspace, times,xcolor}
\usepackage{color, colortbl}
\usepackage{authblk}
\usepackage{arydshln}
\usepackage{braket}
\usepackage{tikz}
\usepackage{verbatim}
\usetikzlibrary{arrows,shapes}
\usepackage{comment}
\usepackage{setspace}
\usepackage[dvips]{epsfig}             % PostScript figures
\usepackage{subfigure}          % subfigures
\usepackage{longtable}          % multipage tables
\usepackage{url, hyperref}                % url?
\usepackage{multirow}           % Multirow table entries
\begin{document}

\title{Quantum Annealing for Prime Factorization}
\date{}
\author[1]{Shuxian Jiang}
\author[2]{Keith A.~Britt}
\author[2]{Alexander J.~McCaskey}
\author[2]{Travis S.~Humble{\thanks{humblets@ornl.gov}}}
\author[1,3]{Sabre Kais{\thanks{kais@purdue.edu}}}
\affil[1]{Department of Computer Science, Purdue University, West Lafayette, IN 47906}
\affil[2]{Quantum Computing Institute, Oak Ridge National Laboratory, Oak Ridge, TN 37831}
\affil[3]{Department of Chemistry, Physics and Birck Nanotechnology Center, Purdue University, West Lafayette, IN 47906}
\maketitle
\begin{abstract}
We have  developed a framework to convert an arbitrary integer factorization problem to an executable Ising model by first writing it as an optimization function and then transforming the k-bit coupling ($k\geq 3$) terms to quadratic terms using ancillary variables. Our resource-efficient method uses  $\mathcal{O}(\text{log}^2(N))$ binary variables (qubits) for finding the factors of an integer $N$. We present how to factorize 15, 143, 59989, and 376289 using 4, 12, 59, and 94 logical qubits, respectively. This method was tested using the D-Wave 2000Q for finding an embedding and determining the prime factors for a given composite number. The method is general and could be used to factor larger integers as the number of available qubits increases.

\end{abstract}

\section{Introduction}
The integer factorization problem reduces an integer $N$ to its prime factors $p$ and $q$. This problem is fundamental in number theory with broad implications for cryptographic data storage and communications \cite{wagstaff2013joy}.  While finding factors of an integer is a computational hard problem, it is not believe to belong to the class of NP-hard problems. In a practical sense though, all known classical factoring algorithms which are deterministic and don't have unproved hypotheses require time exponential in $\log N$.  Thus, the integer factorization problem is used as the basic hardness assumption for many encryption methods including the widely deployed RSA cryptographic system. The fastest, known classical algorithm for integer factorization is the general number field sieve method \cite{lenstra1990number}, which scales exponentially in the number of operations required with respect to the integer $N$. 
\par
Quantum computing theory has shown the potential to reduce the number of operations required for solving certain problems, and many efforts have been undertaken to develop a quantum computer that can solve the integer factorization problem. The quantum methods for solving factorization problem could be regarded as probabilistic methods, compared to the classical deterministic methods. Shor's algorithm is perhaps the most well-known quantum algorithm for integer factorization. To factor an integer $N$, Shor's algorithm requires a polynomial number of operations \cite{shor1999polynomial}, thus providing an exponential speedup over the general number sieve. Shor's algorithm works by reducing the factorization problem to the order-finding problem.  Many attempts have been made to implement Shor's algorithm on quantum computing hardware. Vandersypen et al.~\cite{vandersypen2001experimental} used seven spin-1/2 nuclei in a molecule as qubits to factor $N=15$, while Lanyon et al.~\cite{lanyon2007experimental}, Lu et al.~\cite{lu2007demonstration}, and Politi et al.~\cite{politi2009shor} have implemented compiled versions of Shor's algorithm using photonic systems for factoring $N=15$. Mart{\'\i}n-L{\'o}pez et al.~\cite{martin2012experimental} factored $21$ using qubit recycling, and Lucero et al.~\cite{lucero2012computing} used superconducting qubits to factor 15. Geller et al.\cite{geller2013factoring} used a simplified version of Shor's algorithm for factoring products of the Fermat primes 3, 5, 17, 257, and 65537.
Shor's algorithm is not only useful for solving integer factorization problem, but can be also used to solve other order-finding problems. Very recently, Grosshans et al.\cite{grosshans2015factoring} proposed factoring safe semi-primes using the quantum order finding algorithm which reduced the failure probability. 
\par
Shor's quantum algorithm for order-finding is presented within the circuit model of quantum computation. But another, equally powerful model  of quantum computing (up to a polynomial reduction) is the quantum adiabatic computing model~\cite{farhi2001quantum}(QAC), which can also be used to solve the integer factorization problem. Peng et al.~\cite{peng2008quantum} have developed methods for factoring 21 using QAC and they implemented this algorithm in a three-qubit NMR quantum processor, while Xu et al.~\cite{XZLZPD12} factored 143 using similar NMR technology. A novel approach by Schaller et al.~\cite{schaller2007role} used multiplication tables to produce a set of equations and used those to produce a quadratic cost function. Dridi et al.~\cite{dridi2017prime} further optimized this method using Gröbner bases, reducing the number of auxiliary variables and equations required. 
\par
In this contribution, we introduce a new procedure for solving the integer factorization problem using a variant of QAC known as quantum annealing. Our approach is based on a direct mathematical transformation of the problem to an Ising Hamiltonian, which can be realized using currently available quantum processors.  Next, in order to account for hardware constraints, we introduce a modified multiplication method that reduces the range of the coefficients in the cost function without increasing the number of qubits required. The method is general, resource-efficient, and does not rely on ad-hoc calculations. 
\par
Quantum annealing solves optimization problems using quantum fluctuations \cite{kadowaki1998quantum}. Quantum adiabatic computation (QAC), as developed by Farhi et al. \cite{farhi2001quantum}, approaches the same task given a complex Hamiltonian whose ground state encodes the solution to the optimization problem. This computation begins in the ground state of a simple, well-characterized Hamiltonian, which is then adiabatically evolved to the complex, problem Hamiltonian. According to the adiabatic theorem\cite{M62}, the system state will also evolve the ground state of the problem Hamiltonian provided the evolution is sufficiently slow to prevent excitations to any higher-lying state. At the end of the annealing process, the measured qubits will encode the optimal solution to the problem within a bounded degree of certainty (due to noise within the closed system).
\par
The time-dependent Hamiltonian of the quantum system is given by combining the initial Hamiltonian and the final Hamiltonian~\cite{RevModPhys.90.015002}
\begin{equation}
H(t) = (1- \frac{t}{T})H_B + \frac{t}{T}H_P.
\end{equation} 
Here $H_B$ is the initial Hamiltonian with a well-known and easily constructed ground state, which we consider to have the general form 
\begin{equation}\label{eq:Hb}
H_B=-\sum \sigma_x^{(i)}
\end{equation}
with Pauli operator $\sigma_x$ defining the $x$-basis. The Hamiltonian $H_P$ is the final Hamiltonian whose ground state encodes the solution to a given instance of the optimization problem. Finding the prime factors of an integer will be mapped to the final Ising Hamiltonian of the general form
\begin{equation}
H_P=\sum{{h_i}{\sigma_z^{(i)}}} + \sum{{J_{ij}} }{\sigma_z^{(i)}  }{\sigma_z^{(j)}} 
\end{equation}
where $\sigma_z$ defines the $z$-basis and the local fields ${h_i}$ and the couplings ${J_{ij}}$ define the factorization problem instance. $H_P$ gives the total energy of the system.
\par
The time-dependent Hamiltonian $H(t)$ of the physical system evolves according to Schr\"{o}dinger equation
\begin{equation}\label{eq:Schr}
i\frac{d}{dt}\ket{\psi(t)}=H(t)\ket{\psi(t)}
\end{equation}
where $|\psi(t)\rangle$ is the state of the system at any time $t\in[0,T]$. Let $\ket{\phi_i(t)}$ be the $i$-th instantaneous eigenstate of $H(t)$, that is, $H(t)\ket{\phi_i(t)}=E_i(t)\ket{\phi_i(t)}$ holds through the entire evolution. If the system is initialized in the ground state $|\phi_0(t=0)\rangle$, then 
 the evolution proceeds slow enough to avoid exciting to the higher-lying eigenstates, e.g.,  $|\phi_1(t)\rangle$. Ultimately, the system will be prepared in the instantaneous ground eigenstate $|\phi_0(t=T)\rangle$.
\par
The direct method to factor $N = pq$, where $p$ and $q$ are prime numbers is to let $l_1 = \lfloor \log_2(p) \rfloor, l_2 = \lfloor \log_2(q) \rfloor$ and, without loss of generality, we take $p = (x_{l_1-1}x_{l_1-2}...x_1 1)_2, q = (x_{l_1+l_2-2}x_{l_1+l_2-3}...x_{l_1} 1)_2$, and $l_1>l_2$ where $x_i$ are binary numbers. So $p = \sum_{i=1}^{l_1-1}2^{i}x_i+1$ and $q=\sum_{j=l_1}^{l_1+l_2-2}2^{j}x_j+1$. We can define the cost function $f(x_1,x_2,x_3,x_4,...,x_{l_1+l_2-2})= (N-pq)^2$.
To reduce the order of the cost function to quadratic, 
we need $\binom{l_1}{2} + \binom{l_2}{2} = \frac{l_1(l_1-1)}{2}+ \frac{l_2(l_2-1)}{2}$ auxiliary variables. If $l_1=l_2=l$, the number of auxiliary variables is
$l \times (l-1)$. Plus the variables to denote the factors, we used $ 2\times (l-1) + l \times (l-1) = (l+2) \times (l-1)= \mathcal{O}(l^2)= \mathcal{O}(\text{log}^2(N))$ binary variables in total. We could also let $p = (1 x_{l_1-2}...x_1 1)_2, q = (1 x_{l_1+l_2-4}...x_{l_1-1} 1)_2$ when lengths of $p$ and $q$ are fixed. 
\par
we illustrate this direct method through the factorization of $N = 15$. We have $\log_2(p)\leq 2 <n_2=\log_2(q)<4$, which means $p$ is at most 2 bits, $q$ is at most 3 bits, then we define $p=(x_1 1)_2=x_1\times 2+1, q=(x_2x_3 1)_2=x_2\times 2^2+x_3\times 2+1, x_i \in \{0,1\})$, because $p$ and $q$ are prime numbers. The objective function $f(x_1,x_2,x_3)= (N-pq)^2$ to be minimized has the following form:
\begin{eqnarray*}
f=128x_1x_2x_3-56x_1x_2-48x_1x_3+16x_2x_3-52x_1-52x_2-96x_3+196.
\end{eqnarray*}
Now, we reduced the 3-local term to 2-local term as follows \cite{BH01}:
for $x,y,z \in \{0,1\}$,  $xy = z \ \text{iff} \ xy-2xz-2yz+3z = 0$, and $
xy \neq \ z \ \text{iff} \ xy-2xz-2yz+3z > 0$. It is also easy to check that  
$x_1x_2x_3= x_4x_3+2(x_1x_2-2x_1x_4-2x_2x_4+3x_4)$ if $x_4 = x_1 x_2$, and  
 $x_1x_2x_3 <  x_4x_3+2(x_1x_2-2x_1x_4-2x_2x_4+3x_4)$ if $ \ x_4 \neq x_1 x_2$. Thus, the $x_1x_2x_3$ term could be transformed to quadratic form by replacing $x_1x_2$ with $x_4$, plus a constrained condition as the penalty term:
\begin{eqnarray} \label{eq:replace}
%\min(x_1x_2x_3) =  \underset{x_4=x_1 x_2}{\min}(x_4x_3+2(x_1x_2-2x_1x_4-2x_2x_4+3x_4))
\min(x_1x_2x_3) =  \min(x_4x_3+2(x_1x_2-2x_1x_4-2x_2x_4+3x_4)
\end{eqnarray}	
Here we transformed the 3-local term to 2-local by introducing a new variable and replacing the constrained condition with a penalty term into the original function. We obtain (see Appendix A.1 for more details) an Ising function to be optimized with the local fields $h^T$
and couplings $J$ written as
% \[{\bf h}^T = \bordermatrix{
% 	~ & \sigma_z^{(1)} & \sigma_z^{(2)} & \sigma_z^{(3)} & \sigma_z^{(4)} \cr
% 	~ & 58 & 50 & 12 & -80}\]
% \begin{equation}\label{eq:J}
% \begin{array}{ccl}
% {\bf J} & = & \displaystyle
% \bordermatrix{
% 	~ & \sigma_z^{(1)} & \sigma_z^{(2)} & \sigma_z^{(3)} & \sigma_z^{(4)}   \\[0.02in]
% 	\sigma_z^{(1)} &       & 25    &  -6  &    -64    \\[0.02in]
% 	\sigma_z^{(2)} &       &   &  2   &   -64     \\[0.02in]
% 	\sigma_z^{(3)} &      &    &     & 16    \\[0.02in]
% 	\sigma_z^{(4)} &     &    &    &      \\[0.02in]  }
% \end{array}
% \end{equation}
\begin{figure*}[!htb]
\begin{tabular}{cc}
\hspace{1.0cm}	
\includegraphics[width = 2.3in]{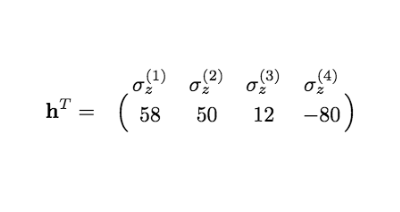}
		& \includegraphics[width = 2.5in]{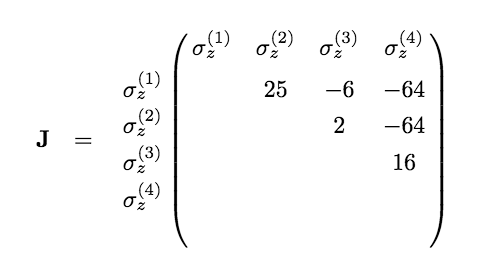}\\
\end{tabular}
\end{figure*}

\subsection{Results}
\label{sec:results}
We use the D-Wave 2000Q quantum annealer to demonstrate the method. %It is necessary to provide the system Hamiltonian in an Ising type model to be solved using D-Wave processor. %It is an important issue to transform the problem Hamiltonian to pairwise interaction as we have shown above.
%There are some techniques\cite{biamonte2008nonperturbative,BH01} to convert the k-qubit($k>2$) interaction terms to 2-qubit terms. In this paper, we provided the factoring problem Hamiltonian and converted it to 2-qubit coupling, so that we could do the experiments using the newly launched D-Wave 2000Q hardware. 
In order to solve problems on the D-Wave hardware we need to embed the problem Hamiltonian onto the \textit{chimera} hardware graph while maintaining the energy minimization objective function \cite{Humble2014}. This process requires users of the D-Wave system to solve two problems: minor embedding\cite{choi2008minor,Klymko2014} and parameter setting\cite{choi2011minor}.
\par
A \emph{minor embedding} of $G(V,E)$ in $G'(V',E')$ is defined by a mapping $\phi:G\mapsto G'$ such that each vertex $v\in G$ is mapped to a connected subtree $T_v$ of $G'$ and if $(u,v)\in E$ then there exist $i_u,i_v\in G'$ such that $i_u\in T_u$, $i_v\in T_v$ and $(i_u,i_v)\in E'$. If such a mapping $\phi$ exists between $G$ and $G'$, we say $G$ is a \emph{minor} of $G'$ and we use $G\le_m G'$ to denote such relationship. 
\par
In parameter setting, we assign each node and each edge in the minor embedding graph such that: (1) for each node in the tree $T_i$ expanded by the same vertex $i$, its value $h'_{i_k}$ satisfies $\sum h'_{i_k} = h_i$, (2) for each edge in the tree $T_i$ expanded by the same vertex $i$, the value $J_{i_{k},i_{k'}}$ needs to be large enough to make sure all physical qubits that correspond to the same logical qubit to be of the same value and (3) for each edge in the minor embedding graph which is in the original graph, we could use the same $J_{ij}$ value. 
 \par
\begin{figure*}[!htb]
\begin{tabular}{cc}
\hspace{-0.2cm}	
\includegraphics[width = 3in]{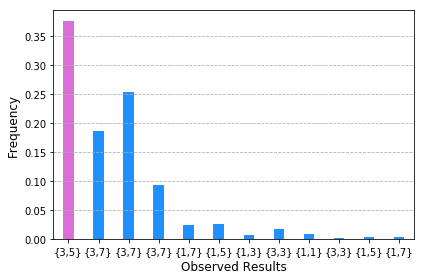}
		& \includegraphics[width = 3in]{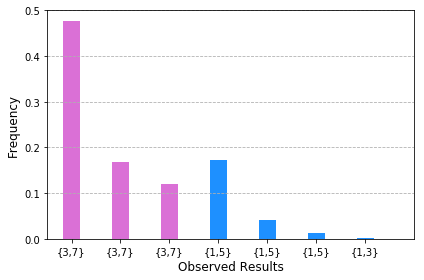}\\
(a)Factoring $15$ with annealing time 200 & (b) Factoring $21$ with annealing time 2000 
\end{tabular}
\caption{Experimental results on D-Wave machine: rates of getting different solutions. For example, the $(3,5)$ in the x-axis denotes the factorization of 15 is 3 multiplied by 5, the number in y-axis denotes the rate to get this factorization. }\label{fig:experiment15}
\end{figure*}

The experimental results of factoring 15 and 21 are shown in Figure \ref{fig:experiment15}. 
These plots show the decoded solutions in order of lowest energy to highest energy (left to right). In some cases, the observed bits were decoded as the correct factors. For example, there are several (3, 7) solutions for $N=21$. Only the first (leftmost) corresponds to the lowest energy state. The others were always higher energy solutions.%The same result (like $\{3,7\}$ in the Figure \ref{fig:experiment15} (a) and (b)) may happen when they differ only on the auxiliary bits.

%\fix{We may change the order of these sections}\\
In order to factor larger numbers and perform the quantum annealing on the D-wave machine with reasonable mitigation of control hardware bits of precision, we introduce the modified multiplication table method. The modified multiplication table method allows us to reduce the range of Ising parameter values used as coefficients to the qubits and couplers, thereby reducing the bits of precision required by control hardware to satisfy the final Hamiltonian.
\par
The modified multiplication table method uses local minimizations of the product of individual bits in the bit strings of $p$ and $q$ to reduce the number of variables needed in the global minimization of the final Hamiltonian. In addition to reducing the number of logical qubits needed to describe the final Hamiltonian, this method also shrinks the range of coefficient values needed to describe the final Hamiltonian. A detailed analysis of the range of coefficients is shown in Appendix A.5. Of note is that the modified multiplication table method did not eliminate the need for 4-body and 3-body terms to be reduced to quadratic terms, but it did reduce the number of these higher body terms making it possible for us to embed them on the D-Wave hardware. We used approximately $\text{log}(N)$ binary variables to denote the factors and about $\text{log}(N)$ binary variables to denote the carries where $N$ is the number to be factored, plus $\text{log}^2(N)/4$ auxiliary binary variables in this scheme. In total, we need roughly $\text{log}^2(N)/4$ binary variables (qubits).
\par
For an illustrative example, we consider factoring $143 = 11 \times 13$. We have the multiplication table\cite{dattani2014quantum}.
\begin{table}[h!]
  \begin{center}
    \caption{Multiplication table for $11\times13=143$ in binary.}
    \label{tab:table1}
    \begin{tabular}{ccccccccc}
      \hline 
\rowcolor{white!85!blue}
& $2^{7}$ & $2^{6}$ & $2^{5}$ & $2^{4}$ & $2^{3}$ & $2^{2}$ & $2^{1}$ & $2^{0}$ \\\hline\hline
\rowcolor{white!98!blue}
p& &      &         &         &   $1$   & $p_{2}$ & $p_{1}$  & $1$ \\
\rowcolor{white!98!blue}
q& &      &         &         &   $1$   & $q_{2}$ & $q_{1}$  & $1$\\\hline
\rowcolor{white!98!blue}
& & & & &\multicolumn{1}{c:}{1} & $p_{2}$& $p_{1}$& \multicolumn{1}{:c}{$1$} \\
\rowcolor{white!98!blue}
& &      &         & \multicolumn{1}{:c}{$q_{1}$}     & \multicolumn{1}{c:}{$p_{2}q_{1}$} & $p_{1}q_{1}$ & $q_{1}$  & \multicolumn{1}{:c}{}   \\
\rowcolor{white!98!blue}
                 &          &          & $q_{2}$  & \multicolumn{1}{:c}{$p_{2}q_{2}$} & \multicolumn{1}{c:}{$p_{1}q_{2}$} & $q_{2}$         &          & \multicolumn{1}{:c}{}   \\
\rowcolor{white!98!blue}
                 &          &   1      & $p_{2}$  & \multicolumn{1}{:c}{$p_{1}$}      & \multicolumn{1}{c:}{1}            &              &          & \multicolumn{1}{:c}{}   \\
\rowcolor{white!85!blue}
carries          &  & $c_{4}$ & $c_{3}$ & \multicolumn{1}{:c}{$c_{2}$}      & \multicolumn{1}{c:}{$c_{1}$}              &      &          & \multicolumn{1}{:c}{}   \\\hline
\rowcolor{white!98!blue}
$p\times q = 143$ & 1        & 0        & 0        &\multicolumn{1}{:c}{0} & 1 & \multicolumn{1}{:c}{1} & 1 & \multicolumn{1}{:c}{1} \\\hline      
\end{tabular}
\end{center}
\end{table}
% \begin{table}[h!]
%   \begin{center}
%     \caption{Multiplication table for $11\times13=143$ in binary.}
%     \label{tab:table1}
%     \begin{tabular}{ccccccccc}
%       \hline 
% & $2^{7}$ & $2^{6}$ & $2^{5}$ & $2^{4}$ & $2^{3}$ & $2^{2}$ & $2^{1}$ & $2^{0}$ \\\hline\hline
% p& &      &         &         &   $1$   & $p_{2}$ & $p_{1}$  & $1$ \\
% q& &      &         &         &   $1$   & $q_{2}$ & $q_{1}$  & $1$\\\hline
%  & &      &         &  \multicolumn{1}{:c}{}&\multicolumn{1}{c:}{1} & $p_{2}$& $p_{1}$& \multicolumn{1}{:c}{$1$} \\
%  & &      &         & \multicolumn{1}{:c}{$q_{1}$}     & \multicolumn{1}{c:}{$p_{2}q_{1}$} & $p_{1}q_{1}$ & $q_{1}$  & \multicolumn{1}{:c}{}   \\
%                   &          &          & $q_{2}$  & \multicolumn{1}{:c}{$p_{2}q_{2}$} & \multicolumn{1}{c:}{$p_{1}q_{2}$} & $q_{2}$         &          & \multicolumn{1}{:c}{}   \\
%                   &          &   1      & $p_{2}$  & \multicolumn{1}{:c}{$p_{1}$}      & \multicolumn{1}{c:}{1}            &              &          & \multicolumn{1}{:c}{}   \\
%  carries          &  & $c_{4}$ & $c_{3}$ & \multicolumn{1}{:c}{$c_{2}$}      & \multicolumn{1}{c:}{$c_{1}$}              &      &          & \multicolumn{1}{:c}{}   \\\hline
% $p\times q = 143$ & 1        & 0        & 0        &\multicolumn{1}{:c}{0} & 1 & \multicolumn{1}{:c}{1} & 1 & \multicolumn{1}{:c}{1} \\\hline      \end{tabular}
%   \end{center}
% \end{table}
Instead of using carries in every bits\cite{dattani2014quantum} \cite{dridi2017prime},\cite{pal2016hybrid}, we only use carries twice (let them to be $c_1,c_2,c_3,c_4$ with $(c_2 c_1)_2=c_2\times 2+c_1$, $(c_4 c_3)_2=c_4\times 2+c_3, c_i = 0,1$) determined by the divided columns. This method skipped the step to calculate the system of equations for each column, thus greatly reduced the burden of computation. To determine how to divide the columns we need to balance the number of unknown variables (carries) and the range of coefficients in the problem Hamiltonian. Instead of making the sum of each column equal to every bits of the number to be factored as in a conventional multiplication table, we make each block of the multiplication table equal to the corresponding block of the number to be factored, thus greatly reducing the number of carry qubits used while keeping their coefficients in a reasonable range (given the hardware platform). The equations for these blocks give the following cost function $f(p_1,p_2,q_1,q_2,c_1,c_2,c_3,c_4)$ to be optimized (more details in Appendix A.2):
\begin{eqnarray*}
 f & = & (2p_2+2p_1 q_1 +2q_2-8c_2-4c_1+p_1+q_1-3)^2+(2q_1+2p_2 q_2 +2p_1+2c_2-8c_4-4c_3+ p_2 q_1+p_1 q_2\\
 & & +c_1+1)^2+(q_2+p_2+c_3+2c_4-2)^2 .
\end{eqnarray*}
This function could be expanded and further simplified using the property $x^2=x \text{ for } x=0,1$. But there will still be cubic terms like $c_1p_1q_1$ and quartic terms like $p_1p_2q_1q_2$. In order to convert it to Ising Hamiltonian, we need to reduce these high order terms to two order terms as explained
in the Appendix A.2.  Replacing $p_1q_1$ with $t_1$, $p_1q_2$ with $t_2$, $p_2q_2$ with $t_3$, and $p_2q_1$ with $t_4$, we used $2\times 2 =4$ auxiliary variables. After doing a variable conversion from $p_1$ to $s_1$, $p_2$ to $s_2$, $q_1$ to $s_3$, $q_2$ to $s_4$, ..., $p_2q_1$ to $s_{12}$ as the final step, we transfer the cost function to an Ising type Hamiltonian as shown in Appendix A.2.
\par
\begin{figure}[htb!] \label{fig: chimera_12}\center
 \includegraphics[width=0.6\linewidth]{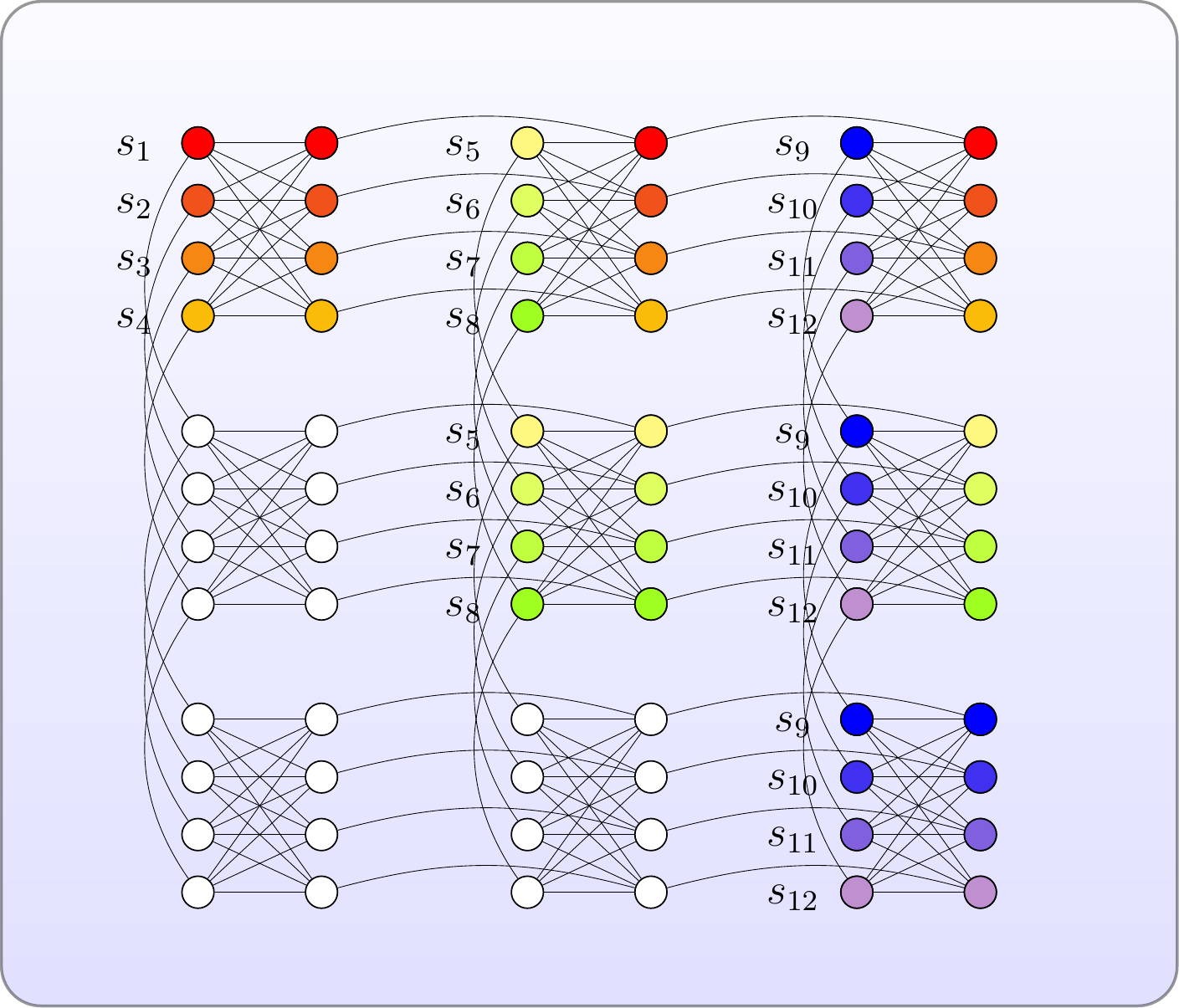}\center
 \caption{Embedding the factoring instance $N=143$ to Chimera graph. The nodes with the same color denote the same original qubit, with their connected lines corresponding to strong couplings. The left footnotes refer to which spin the node was embedded.} \label{emb_143}
 \end{figure}

Next we embed the problem to D-wave machine using the following method (note: for larger $N$ factoring experiments which can't be embedded directly using this method due to the limitation of the Chimera graph, we relied on a D-Wave provided heuristic embedding algorithm). If $n$ qubits are needed in the Hamiltonian, we divide $n$ into $\lceil \frac{n}{4} \rceil$ groups. For each group, we use 4 copies of the nodes with each $h'_{i_k} = \frac{1}{4} h_i$. We assign each edge in the tree $T_i$ the negative number with largest absolute value to make it a penalty term. This method guarantees the nodes correspond to the same original qubit have the same value. We assign each edge corresponds to the original edge in the problem graph the same $J_{ij}$ value. The embedded graph to D-Wave machine is in Figure \ref{emb_143}.

\begin{figure*}[!htb]
\center
\begin{tabular}{cc}	
\includegraphics[width = 2.6in]{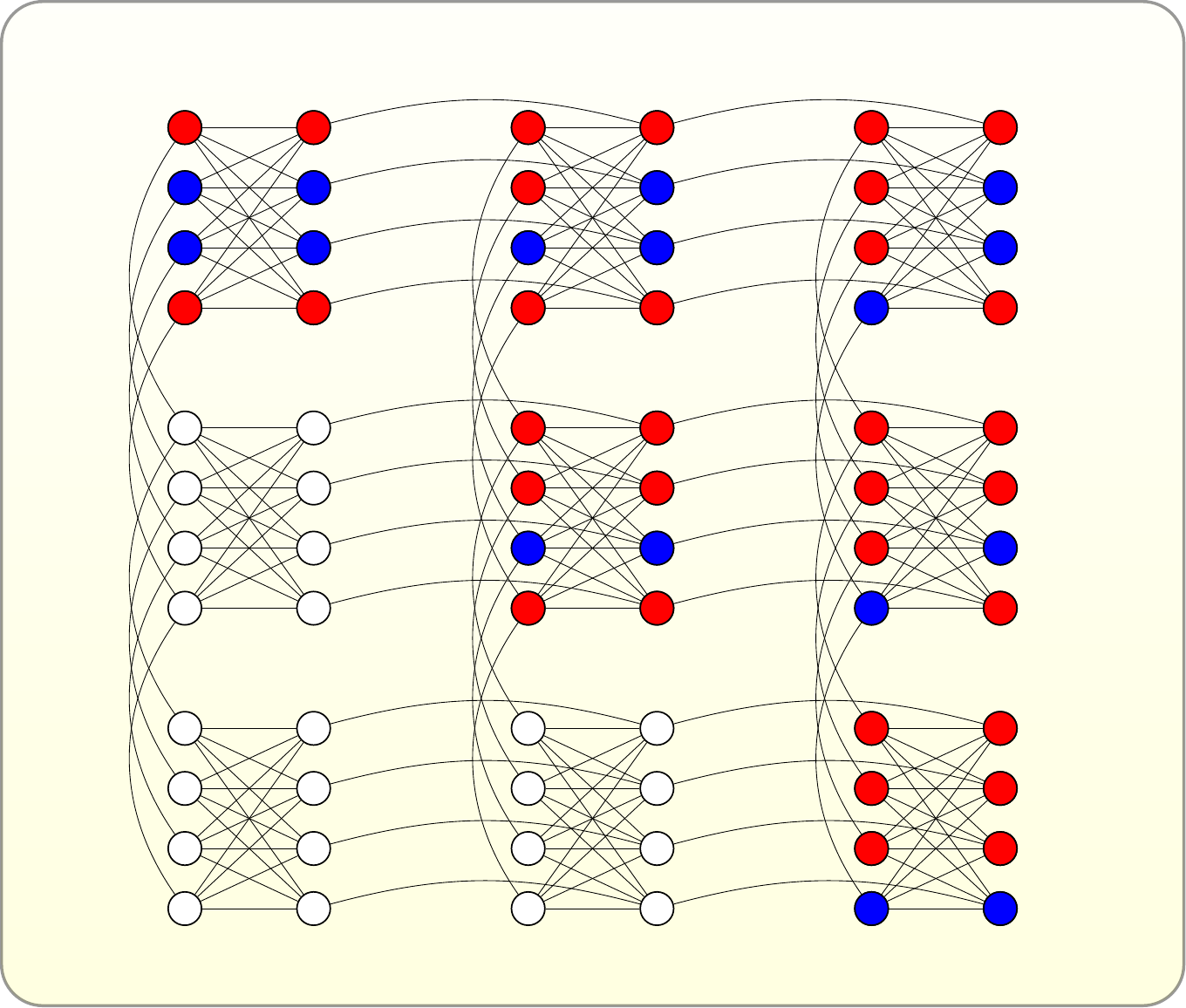}
		& \includegraphics[width = 2.6in]{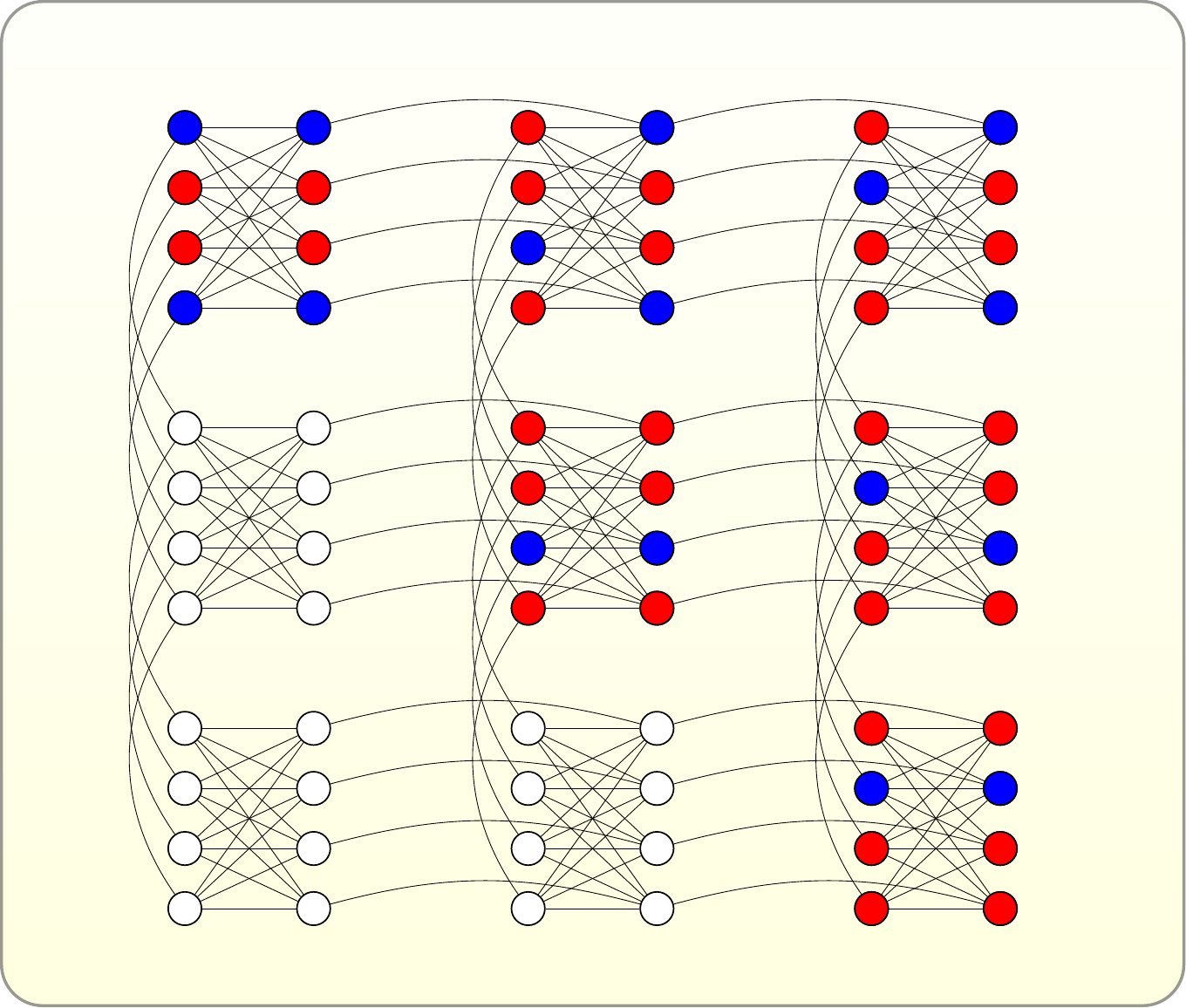}\\
(a) Result $13 \times 11$ & (b) Result $11 \times 13$
\end{tabular}
\caption{Experimental results on D-wave machine: final ground state of factoring 143. Nodes colored red denote $+1$, nodes colored blue denote $-1$. (a) This graph shows $s_1=1, s_2=-1, s_3=-1, s_4=1$ which means $p=1101, q= 1011$. (b) This graph shows $s_1=-1, s_2=1, s_3=1, s_4=-1$ which means $p = 1011, q=1101$.}\label{fig:results 143}
\end{figure*}

%To complete the simulations, we prepare the initial  state to be the ground state of $H_B=-\sum_{i=1}^n \sigma_x^{(i)}$ 
%(2). Assign the problem Hamiltonian in the form of
%$H_P=\sum_{i=1}^n h_i\sigma_z^{(i)}+\sum_{i,j}^n {J_{ij}\sigma_z^{(i)}\sigma_z^{(j)}}$ %using \textsc{Ising}$({\bf h},{\bf J})$ defined above.   
%(3). As system evolves adiabaticly according to Hamiltonian $H(t)$ , the final state $\ket{\psi(t=T)}$ will stay at the ground state. After a measurement, 
The results graph are shown in Figure \ref{fig:results 143}. The final state of the system will be  $\ket{1 -1 -1 \ 1}$ or $\ket{-1 \ 1 \ 1 -1}$ with high probability, which corresponds to solutions $p=(1 p_2 p_1 1)=(1 1 0 1)_2 = 13$, $q=(1 q_2 q_1 1) = (1 0 1 1)_2 = 11$ or $p = 11, q= 13$.

Next, we factorize larger numbers such as $N = 59989 = pq$, where $p>q$ are prime numbers. We start by fixing  the length of the factors by setting the binary representation of the factors to $p = (1 p_6 p_5 p_4 p_3 p_2 p_1 1)_2,q = (1 q_6 q_5 q_4 q_3 q_2 q_1 1)_2$, $p_i, q_i\in \{0,1\}$. We predefine how to divide the columns and the number of carries given the multiplication table shown in Appendix A.3. Using the method described for factoring 143, we write the equations and the corresponding cost function of the factorization problem then convert this cost function to the Ising Hamiltonian. We need $6\times 6 = 36$ auxiliary variables, based on the same variable replacement rules stated in the previous section, plus $6 + 6= 12$ variables to denote the factors and 11 variables to denote the carries. This gives a total number of 59 variables. Sometimes the carries in the multiplication table will overlap, as is the case for factoring $376289 = 659\times 571$ shown in the multiplication table of Appendix A.4. In such circumstances, we add these carries in the table and then use the same method as before to find the corresponding Ising Hamiltonian. This Hamiltonian has $8+8+14+8\times8 =94$ qubits. As a point of reference, applying this method to the current factoring record for RSA-768 would require approximately 147,456 qubits. %$382+382+382*382+768$ 
\par
%We embed this Hamiltonian into the D-wave Chimera hardware using the embedding function \textit{find\_embedding} and \textit{embed\_problem} available in the vendor's software developer packages for Python. Based on the embedded result obtained from these functions, the total number of qubits required using the Chimera graphs is 1,070. This number may be further improved depending on the performance of the predefined functions, which are based on heuristic, random search methods, or by designing specialized embeddings of this Hamiltonian for the Chimera graph. See \cite{cai2014practical}\cite{boothby2016fast} for research related to such work. %Finally, we have used the functions \textit{solve\_ising} and \textit{unembed\_answer} to recover solutions to these problem using the D-Wave quantum annealing hardware.

\section{Conclusions}
In this paper we have presented two general methods %with relatively inexpensive preprocessing 
for factoring integers using quantum annealing by converting the problem to an Ising Hamiltonian. Both methods use $\mathcal{O}(\text{log}^2(N))$ qubits in total, where $N$ is the number to be factored. The novelty of our demonstration of quantum annealing for prime factorization is based on the reduction in quantum resources required to execute factoring and the experimental verification of the algorithmic accuracy using currently available hardware.  As a proof-of-concept, we have demonstrated these methods by factoring integers using the D-Wave 2000Q quantum annealing hardware, but these methods may be used on any other quantum annealing system with a similar number of qubits, qubit degree of connectivity, and hardware parameter precision. Assuming that quantum annealing hardware systems will continue to grow both in the number of qubits and bits of precision capabilities, our methods offer a promising path toward factor much larger numbers in the future.
\par
Finally, we note that while our demonstrations of factoring have made use of currently available quantum annealers, there is an outstanding question regarding the asymptotic complexity for this approach. It is well known that algorithmic complexity within the QAC model depends on the minimum spectral gap between the ground and first-excited states of the underlying time-dependent Hamiltonian. Attempts to classify the complexity of the spectral gap with respect to system size have not yet succeed and, indeed, Cubitt, Perez-Garcia, and Wolf have proven that the problem of claiming a Hamiltonian has a gap is undecidable in general \cite{cubitt2015undecidability}. Nonetheless, there is hope that our resource-efficient algorithms may find use in pre-processing potential factors for noisy factorization algorithms, e.g., as suggested by Patterson et al.~within the context of RSA~\cite{Patterson2012}.

{\bf {Methods}}
%\fix{Do we move this part to sec. 1.1?}\\
We calculated the factors $p$ and $q$ of a coprime integer $N$ using a implementation of the algorithms described in Sec.~\ref{sec:results}. The programmed implementation was written in C/C++ or Python using the XACC programming framework \cite{mccaskey2017extreme}. XACC enables integration of the D-Wave solver application programming interface (SAPI) using a directive-based programming model. Pre-processing of the input $N$ generated the Ising parameters for a logical Hamiltonian that was then embedded into the hardware graph structure. For the 2000Q processor, the hardware graph was a complete 16-by-16 Chimera structure over 2048 qubits using the SAPI version 3.0 sapi\_findembedding method, which is based on the Cai, Macready and Roy randomized algorithm \cite{cai2014practical}. Access to these methods were managed using the XACC dwsapi-embedding plugin \cite{mccaskey2017extreme}. The corresponding biases and couplings for the embedded problem were generated using the logical Ising parameters. The output of the embeddeding was a program implementation of the physical Ising model that was submitted for execution on the D-Wave processor. Additional parameters for the execution included the number of samples $S$ and the annealing duration $T$. The default annealing schedule for the 2000Q was used for all executions. The output from each of the $S$ executions was a measured binary string designating $\pm1$ values for each spin variable. The number of samples was $S = 10,000$. Each returned string was then classified according to the corresponding energy for the physical Ising model and subsequently decoded according to the algorithm in Sec.~\ref{sec:results} into the factors $p$ and $q$. A histogram of all solutions returned for a specific annealing time was recorded. 

{\bf Acknowledgments}
Access to the D-Wave 2000Q was provided by the Quantum Computing Institute at Oak Ridge National Laboratory and Google Quantum Artificial Intelligence Lab, USRA-Purdue. This manuscript has been authored by UT-Battelle, LLC under Contract No. DE-AC05-00OR22725 with the U.S. Department of Energy. The United States Government retains and the publisher, by accepting the article for publication, acknowledges that the United States Government retains a non-exclusive, paid-up, irrevocable, worldwide license to publish or reproduce the published form of this manuscript, or allow others to do so, for United States Government purposes. The Department of Energy will provide public access to these results of federally sponsored research in accordance with the DOE Public Access Plan (http://energy.gov/downloads/doe-public-access-plan).

\bibliographystyle{plain}
{\small
\bibliography{ref}
}
\appendix
\section{Appendix}

\subsection{Factoring $N = 15 = 5 \times 3$}
Define $p=(x_1 1)_2=x_1*2+1, q=(x_2x_3 1)_2=x_2*2^2+x_3*2+1, x_i \in \{0,1\})$, $p$ and $q$ are prime numbers. \\ 
The cost function is 
\begin{eqnarray*}
 & & f(x_1,x_2,x_3) \\
& = & (N-pq)^2 \\
& = & [15-(x_1*2+1)(x_3*2^2+x_2*2+1)]^2 \\
%& = & 15^2+(2x_1+1)^2(4x_3+2x_2+1)^2-30(2x_1+1)(4x_3+2x_2+1) \\
& = & 128x_1x_2x_3-56x_1x_2-48x_1x_3+16x_2x_3-52x_1-52x_2-96x_3+196.
\end{eqnarray*}
Use the replacement in Eq.\ref{eq:replace}, we got
 \begin{eqnarray*}
 & & f'(x_1,x_2,x_3,x_4) \\
 & = & 128(x_4x_3+2(x_1x_2-2x_1x_4-2x_2x_4+3x_4))-56x_1x_2-48x_1x_3+16x_2x_3\\
 & & -52x_1-52x_2-96x_3+196 \\
 & = &200x_1x_2 - 48x_1x_3 - 512x_1x_4 + 16x_2x_3 - 512x_2x_4 + 128x_3x_4 \\
 & & - 52x_1- 52x_2 - 96x_3+768x_4+ 196.
\end{eqnarray*}
with 
\begin{equation*}
 \underset{x_1 x_2=x_4} {\min} f(x_1,x_2,x_3,x_4) = {\min}f'(x_1,x_2,x_3,x_4) 
\end{equation*}
because the coefficient of $x_1x_2x_3$ term is positive.

Then we do variable replacement using $x_i= \frac{1-s_i}{2}, i = 1,2,3,4$
\begin{eqnarray*}
& & f'(x_1,x_2,x_3,x_4) \\
& = & 200\frac{1-s_1}{2}\frac{1-s_2}{2} -48\frac{1-s_1}{2}\frac{1-s_3}{2}-512\frac{1-s_1}{2}\frac{1-s_4}{2}+16\frac{1-s_2}{2}\frac{1-s_3}{2}\\
& & -512\frac{1-s_2}{2}\frac{1-s_4}{2}+128\frac{1-s_3}{2}\frac{1-s_4}{2}-52\frac{1-s_1}{2}-52\frac{1-s_2}{2}\\
& & -96\frac{1-s_3}{2}+768\frac{1-s_4}{2}+196\\
%& = & 50(1-s_1)(1-s_2) -12(1-s_1)(1-s_3)-128(1-s_1)(1-s_4)\\
%& & +4(1-s_2)(1-s_3) -128(1-s_2)(1-s_4) +32(1-s_3)(1-s_4)\\
%& & -26(1-s_1)-26(1-s_2)-48(1-s_3)+384(1-s_4)+196 \\
& = & 116s_1 + 100s_2 + 24s_3 - 160s_4 + 50s_1s_2 -12s_1s_3 - 128s_1s_4 + 4s_2s_3 -128s_2s_4 +32s_3s_4 + 298\\
& = & 2g(s_1,s_2,s_3,s_4)
\end{eqnarray*}
$g(s_1,s_2,s_3,s_4)$ is the energy function of  
\begin{eqnarray*}
& & H_P  (\sigma_z^{(1)},\sigma_z^{(2)},\sigma_z^{(3)},\sigma_z^{(4)})\\
& =&58\sigma_z^{(1)}+50\sigma_z^{(2)}+12\sigma_z^{(3)}-80\sigma_z^{(4)}+25\sigma_z^{(1)}\sigma_z^{(2)}-6\sigma_z^{(1)}\sigma_z^{(3)}-64 \sigma_z^{(1)}\sigma_z^{(4)}+2\sigma_z^{(2)}\sigma_z^{(3)}-64\sigma_z^{(2)}\sigma_z^{(4)}\\
& & +16\sigma_z^{(3)}\sigma_z^{(4)}+149I. 
\end{eqnarray*}	

\subsection{Factoring $N = 143 = 13 \times 11$}
From multiplication table \ref{tab:table1}, we could get the equations for each blocks
\begin{eqnarray*}
 (p_2+p_1 q_1 +q_2-(c_2\times 4 + c_1\times 2))\times 2+(p_1+q_1) &=& (1 1)_2 = 3 \\
 (q_1+p_2 q_2 +p_1 + c_2-(c_4\times 4 +c_3\times 2))\times 2 + (1+p_2 q_1+p_1 q_2 +1+ c_1)& = & (0 1)_2=1\\
(1+c_4)\times 2+ (q_2+p_2+c_3)& = & (1 0 0)_2=4
\end{eqnarray*}
They could be further simplified as 
\begin{eqnarray*}
2p_2+2p_1 q_1 +2q_2-8c_2-4c_1+p_1+q_1-3 &=& 0 \\
2q_1+2p_2 q_2 +2p_1+2c_2-8c_4-4c_3+ p_2 q_1+p_1 q_2+c_1+1& = & 0 \\
q_2+p_2+c_3+2c_4-2& = & 0
\end{eqnarray*}
We define the cost function to be squares of the left of equations. That is
\begin{eqnarray*}
  & & f(p_1,p_2,q_1,q_2,c_1,c_2,c_3,c_4) \\
  & = & (2p_2+2p_1 q_1 +2q_2-8c_2-4c_1+p_1+q_1-3)^2+(2q_1+2p_2 q_2 +2p_1+2c_2-8c_4-4c_3+ p_2 q_1+p_1 q_2+c_1+1)^2\\
  & & +(q_2+p_2+c_3+2c_4-2)^2
\end{eqnarray*}
Expand and simplify the function using the property $x^2=x \text{ for } x=0,1$. Then reduce the higher order terms to two order terms according to the following rule noticing that there will be negative high order terms:
\[
 \begin{cases}
x_1x_2x_3 = x_4x_3+2(x_1x_2-2x_1x_4-2x_2x_4+3x_4) \ \  \text{if} \ x_4 = x_1 x_2\\
x_1x_2x_3 < x_4x_3+2(x_1x_2-2x_1x_4-2x_2x_4+3x_4) \ \  \text{if} \ x_4 \neq x_1 x_2 
  \end{cases}
\]
\[
and \begin{cases}
-x_1x_2x_3 = -x_4x_3+2(x_1x_2-2x_1x_4-2x_2x_4+3x_4) \ \  \text{if} \ x_4 = x_1 x_2\\
 -x_1x_2x_3 < -x_4x_3+2(x_1x_2-2x_1x_4-2x_2x_4+3x_4) \ \  \text{if} \ x_4 \neq x_1 x_2
  \end{cases}
\]
%\begin{eqnarray}
% x_1x_2x_3 & = & x_4x_3+2(x_1x_2-2x_1x_4-2x_2x_4+3x_4) \ \  \text{if} \ x_4 = x_1 x_2\\
%  x_1x_2x_3 & < & x_4x_3+2(x_1x_2-2x_1x_4-2x_2x_4+3x_4) \ \  \text{if} \ x_4 \neq x_1 x_2 \\
%  -x_1x_2x_3 & = & -x_4x_3+2(x_1x_2-2x_1x_4-2x_2x_4+3x_4) \ \  \text{if} \ x_4 = x_1 x_2\\
%  -x_1x_2x_3 & < & -x_4x_3+2(x_1x_2-2x_1x_4-2x_2x_4+3x_4) \ \  \text{if} \ x_4 \neq x_1 x_2
% \end{eqnarray}

So the negative term $-x_1x_2x_3$ could be transformed to quadratic term in the same way as the positive term $x_1x_2x_3$.

The cost function could be minimized as long as the transformed one is minimized 
\begin{eqnarray} \label{eq:replace}
%\min(x_1x_2x_3) & = &   \underset{x_4=x_1 x_2}{\min}(x_4x_3+2(x_1x_2-2x_1x_4-2x_2x_4+3x_4)) \\
%\min(-x_1x_2x_3) & = &   \underset{x_4=x_1 x_2}{\min}(-x_4x_3+2(x_1x_2-2x_1x_4-2x_2x_4+3x_4))
\min(x_1x_2x_3) & = &   \min(x_4x_3+2(x_1x_2-2x_1x_4-2x_2x_4+3x_4) \\
\min(-x_1x_2x_3) & = & \min(-x_4x_3+2(x_1x_2-2x_1x_4-2x_2x_4+3x_4)
\end{eqnarray}	

Replace $p_1q_1$ with $t_1$,$p_1q_2$ with $t_2$, $p_2q_2$ with $t_3$, $p_2q_1$ with $t_4$, using the variable replacement rule if the coefficient of the term is positive or negative respectively. The cost function becomes
\begin{eqnarray*}
   & & f(p_1,p_2,q_1,q_2,c_1,c_2,c_3,c_4,t_1,t_2,t_3,t_4) \\
%  & = & 43c_1 + 120c_2 + 5c_3+ 44c_4 + 3p_1 - 11p_2 + 3q_1 - 11q_2 + 68c_1c_2 - 8c_1c_3 - 16c_1c_4 - 16c_2c_3 - 32c_2c_4 + 68c_3c_4 \\
%  & &  - 4c_1p_1- 16c_1p_2 - 8c_2p_1- 32c_2p_2 - 16c_3p_1 + 2c_3p_2 - 32c_4p_1 + 4c_4p_2 - 4c_1q_1- 16c_1q_2 - 8c_2q_1- 32c_2q_2 \\
%  & & - 16c_3q_1 + 2c_3q_2 - 32c_4q_1 + 4c_4q_2 + 4p_1p_2 + 10p_1q_1 + 11p_1q_2 + 11p_2q_1 + 18p_2q_2 + 4q_1q_2 \\
%  & & +(12p_2+12q_2)p_1q_1 + (- 16c_1- 32c_2)p_1q_1 \\
%  & & + (2c_1+ 4c_2+12p_2)p_1q_2 +(- 8c_3- 16c_4)p_1q_2\\
%  & & + (2c_1+ 4c_2+12q_2)p_2q_1+(- 8c_3- 16c_4)p_2q_1 \\
%  & & + (4c_1+ 8c_2)p_2q_2 + (- 16c_3- 32c_4)p_2q_2\\
%  & & + 2p_1q_1p_2q_2 + 14\\
%  & = & 43c_1 + 120c_2 + 5c_3+ 44c_4 + 3p_1 - 11p_2 + 3q_1 - 11q_2 + 68c_1c_2 - 8c_1c_3 - 16c_1c_4 - 16c_2c_3 - 32c_2c_4 + 68c_3c_4 \\
%  & &  - 4c_1p_1- 16c_1p_2 - 8c_2p_1- 32c_2p_2 - 16c_3p_1 + 2c_3p_2 - 32c_4p_1 + 4c_4p_2 - 4c_1q_1- 16c_1q_2 - 8c_2q_1- 32c_2q_2 \\
%  & & - 16c_3q_1 + 2c_3q_2 - 32c_4q_1 + 4c_4q_2 + 4p_1p_2 + 10p_1q_1 + 11p_1q_2 + 11p_2q_1 + 18p_2q_2 + 4q_1q_2 \\
%  & & +(12p_2+12q_2)t_1 + (- 16c_1- 32c_2)t_1+(24+48)*2(p_1q_1-2p_1t_1-2q_1t_1+3t_1) \\
%  & & + (2c_1+ 4c_2+12p_2)t_2 +(- 8c_3- 16c_4)t_2+(18+24)*2(p_1q_2-2p_1t_2-2q_2t_2+3t_2)\\
%  & & + (2c_1+ 4c_2+12q_2)t_4+(- 8c_3- 16c_4)t_4+(18+24)*2(p_2q_1-2p_2t_4-2q_1t_4+3t_4) \\
%  & & + (4c_1+ 8c_2)t_3 + (- 16c_3- 32c_4)t_3+(12+48)*2(p_2q_2-2p_2t_3-2q_2t_3+3t_3)\\
%  & & + 2t_1t_3 +2*2(p_1q_1-2p_1t_1-2q_1t_1+3t_1)+2*2(p_2q_2-2p_2t_3-2q_2t_3+3t_3)+ 14\\
 & = & 43 c_1 + 120 c_2 + 5 c_3 + 44 c_4 + 3 p_1 - 11 p_2 + 3 q_1 - 11 q_2 + 444 t_1 + 252 t_2 + 372 t_3 + 252 t_4 + 68 c_1 c_2 - 8 c_1 c_3 \\
  & & - 16 c_1 c_4 - 16 c_2 c_3 - 32 c_2 c_4 + 68 c_3 c_4 - 4 c_1 p_1 - 16 c_1 p_2 - 8 c_2 p_1 - 32 c_2 p_2 - 16 c_3 p_1 + 2 c_3 p_2 - 32 c_4 p_1 + 4 c_4 p_2 \\
  & & - 4 c_1 q_1 - 16 c_1 q_2 - 8 c_2 q_1 - 32 c_2 q_2 - 16 c_3 q_1 + 2 c_3 q_2 - 32 c_4 q_1 + 4 c_4 q_2 - 16 c_1 t_1 + 2 c_1 t_2 - 32 c_2 t_1 + 4 c_1 t_3\\
 & & + 4 c_2 t_2 + 2 c_1 t_4 + 8 c_2 t_3 - 8 c_3 t_2 + 4 c_2 t_4 - 16 c_3 t_3 - 16 c_4 t_2 - 8 c_3 t_4 - 32 c_4 t_3 - 16 c_4 t_4 + 4 p_1 p_2 + 158 p_1 q_1 \\
  & & + 95 p_1 q_2 + 95 p_2 q_1 + 142 p_2 q_2 + 4 q_1 q_2 - 296 p_1 t_1 - 168 p_1 t_2 + 12 p_2 t_1 + 12 p_2 t_2 - 248 p_2 t_3 - 168 p_2 t_4 - 296 q_1 t_1 \\
 & & + 12 q_2 t_1 - 168 q_2 t_2 - 168 q_1 t_4 - 248 q_2 t_3 + 12 q_2 t_4 + 2 t_1 t_3 + 14
\end{eqnarray*}
Then we do a variable transformation to make the variable in the domain \{-1,1\} using $x_i=\frac{1-s_i}{2}$ if we let $x_1=p_1, x_2=p_2,..., x_{12}=t_4$.
\begin{table}[h!]
  \begin{center}
    \begin{tabular}{ccccccccccccc}
& $p_1$ & $p_2$ & $q_1$ & $q_2$ & $c_1$ & $c_2$ & $c_3$ & $c_4$ & $t_1$ & $t_2$ & $t_3$ & $t_4$ \\
& $\downarrow$ & $\downarrow$  & $\downarrow$  & $\downarrow$  & $\downarrow$  & $\downarrow$ & $\downarrow$  & $\downarrow$  & $\downarrow$  & $\downarrow$  & $\downarrow$  & $\downarrow$  \\ 
& $s_1$ & $s_2$ & $s_3$ & $s_4$ & $s_5$ & $s_6$ & $s_7$ & $s_8$ & $s_9$ & $s_{10}$ & $s_{11}$ & $s_{12}$
  \end{tabular}
  \end{center}
\end{table}
\begin{eqnarray*}
  & & f'(p_1,p_2,q_1,q_2,c_1,c_2,c_3,c_4,t_1,t_2,t_3,t_4) \\
   & = & 2f(p_1,p_2,q_1,q_2,c_1,c_2,c_3,c_4,t_1,t_2,t_3,t_4) \\
%   & = & 43(1-s_5) + 120(1-s_6) + 5(1-s_7)+ 44(1-s_8)+ 3(1-s_1) - 11(1-s_2) + 3(1-s_3) - 11(1-s_4)\\
%   & & +444(1-s_9)+372(1-s_{11})+252(1-s_{10})+252(1-s_{12})+ 34(1-s_5)(1-s_6) - 4(1-s_5)(1-s_7) \\
%   & & - 8(1-s_5)(1-s_8) - 8(1-s_6)(1-s_7) - 16(1-s_6)(1-s_8) + 34(1-s_7)(1-s_8) - 2(1-s_5)(1-s_1) \\
%   & & - 8(1-s_5)(1-s_2) -4(1-s_6)(1-s_1)-16(1-s_6)(1-s_2) - 8(1-s_7)(1-s_1) + (1-s_7)(1-s_2) \\
%   & & - 16(1-s_8)(1-s_1) + 2(1-s_8)(1-s_2)- 2(1-s_5)(1-s_3)- 8(1-s_5)(1-s_4)- 4(1-s_6)(1-s_3) \\
%   & & - 16(1-s_6)(1-s_4) - 8(1-s_7)(1-s_3) + (1-s_7)(1-s_4) - 16(1-s_8)(1-s_3)+ 2(1-s_8)(1-s_4) \\
%   & & + 2(1-s_1)(1-s_2) +79(1-s_1)(1-s_3) +47.5(1-s_1)(1-s_4) + 47.5(1-s_2)(1-s_3) +71 (1-s_2)(1-s_4) \\
%   & & + 2(1-s_3)(1-s_4)-8(1-s_5)(1-s_9)+(1-s_5)(1-s_{10})-16(1-s_6)(1-s_9)-148(1-s_1)(1-s_9) \\
%   & & -84(1-s_1)(1-s_{10})+6(1-s_2)(1-s_9)+6(1-s_2)(1-s_{10})-124(1-s_2)(1-s_{11})-84(1-s_2)(1-s_{12}) \\
%   & & -148(1-s_3)(1-s_9)-84(1-s_3)(1-s_{12})+6(1-s_4)(1-s_9)-84(1-s_4)(1-s_{10}) -124(1-s_4)(1-s_{11}) \\
%   & & +6(1-s_4)(1-s_{12})+ 2(1-s_5)(1-s_{11})+ (1-s_5)(1-s_{12})+2(1-s_6)(1-s_{10})+4(1-s_6)(1-s_{11})\\
%   & & +2(1-s_6)(1-s_{12})-4(1-s_7)(1-s_{10})-8(1-s_7)(1-s_{11})-4(1-s_7)(1-s_{12})-8(1-s_8)(1-s_{10})\\
%   & & -16(1-s_8)(1-s_{11})-8(1-s_8)(1-s_{12})+(1-s_9)(1-s_{11})+14\\
 & = & (261s_1)/2 + (215s_2)/2 + (261s_3)/2 + (215s_4)/2 - 41s_5 - 82s_6 + 3s_7 + 6s_8 - 137s_9 - 81s_{10} - 107s_{11} - 81s_{12} \\
 & & + 2s_1s_2 + 79s_1s_3 + (95s_1s_4)/2 + (95s_2s_3)/2 - 2s_1s_5 + 71s_2s_4 - 4s_1s_6 - 8s_2s_5 + 2s_3s_4 - 8s_1s_7 - 16s_2s_6 \\
 & & - 2s_3s_5 - 16s_1s_8 + s_2s_7 - 4s_3s_6 - 8s_4s_5 - 148s_1s_9 + 2s_2s_8 - 8s_3s_7 - 16s_4s_6 - 84s_1s_{10} + 6s_2s_9 - 16s_3s_8 \\
& & + s_4s_7 + 34s_5s_6 + 6s_2s_{10} - 148s_3s_9 + 2s_4s_8 - 4s_5s_7 - 124s_2s_{11} + 6s_4s_9 - 8s_5s_8 - 8s_6s_7 - 84s_2s_{12} - 84s_4s_{10} \\
& & - 8s_5s_9 - 16s_6s_8 - 84s_3s_{12} - 124s_4s_{11} + s_5s_{10} - 16s_6s_9 + 34s_7s_8 + 6s_4s_{12} + 2s_5s_{11} + 2s_6s_{10} + s_5s_{12} + 4s_6s_{11} \\
 & & - 4s_7s_{10} + 2s_6s_{12} - 8s_7s_{11} - 8s_8s_{10} - 4s_7s_{12} - 16s_8s_{11} - 8s_8s_{12} + s_9s_{11} + 794
\end{eqnarray*}
This corresponds to Ising Hamiltonian with local fields 
\[{\bf h}^T = \bordermatrix{
	~ & \sigma_z^{(1)} & \sigma_z^{(2)} & \sigma_z^{(3)} & \sigma_z^{(4)} & \sigma_z^{(5)} & \sigma_z^{(6)} & \sigma_z^{(7)} & \sigma_z^{(8)} & \sigma_z^{(9)} & \sigma_z^{(10)} & \sigma_z^{(11)} & \sigma_z^{(12)} \cr
	~ & 130.5 & 107.5 & 130.5 &107.5&-41&-82&3&6&-137&-81&-107&-81}\]
and coupling terms:
\begin{equation*}\label{eq:J}
\begin{array}{ccl}
{\bf J} & = & \displaystyle
\bordermatrix{
	~ & \sigma_z^{(1)} & \sigma_z^{(2)} & \sigma_z^{(3)} & \sigma_z^{(4)}  & \sigma_z^{(5)} & \sigma_z^{(6)}& \sigma_z^{(7)}& \sigma_z^{(8)}& \sigma_z^{(9)}& \sigma_z^{(10)}& \sigma_z^{(11)}& \sigma_z^{(12)} \\[0.02in]
	\sigma_z^{(1)} & &2&79&47.5&-2&-4& -8 & -16 &-148 &-84 &0 & 0\\[0.02in]
	\sigma_z^{(2)} & & &47.5&71& -8 & -16 & 1 & 2 & 6 &6 &-124&-84 \\[0.02in]
	\sigma_z^{(3)} & & & & 2 &-2&-4& -8 & -16 &-148&0 &0 & -84\\[0.02in]
	\sigma_z^{(4)} & & & & & -8 & -16 & 1 & 2 & 6& -84 & -124 & 6\\[0.02in] 
    \sigma_z^{(5)} & & & & & & 34 & -4 & -8 & -8 & 1& 2& 1 \\[0.02in] 
    \sigma_z^{(6)} & & & & & & & -8 & -16 & -16 & 2& 4& 2\\[0.02in] 
    \sigma_z^{(7)} & & & & & & & & 34 & 0 & -4 &-8 & -4 \\[0.02in]
    \sigma_z^{(8)} & & & & & & & & & 0& -8 & -16 & -8    \\[0.02in]
    \sigma_z^{(9)} & & & & & & & & & &0 & 1 &  0        \\[0.02in]
    \sigma_z^{(10)}& & & & & & & & & & & 0  & 0      \\[0.02in]
    \sigma_z^{(11)}& & & & & & & & & & &   & 0      \\[0.02in]
    \sigma_z^{(12)}& & & & & & & & & & &   &       \\[0.02in]}
\end{array}
\end{equation*}

\subsection{Factoring $ N = 59989 = 251\times 239$ }
The following table shows how to divide the columns to blocks. The lengths of each carries ($c_{11}c_{10}c_9$, $c_8c_7c_6$, $c_5c_4c_3$, $c_2c_1$) are determined by what is the largest carry for the numbers in right next block are(assuming each variable to be 1, then add them up). For example, the maximum carry for the right-most block (except the least significant bit) is 3 which is 11 in binary, so the length of the carry for this block is 2. This carry is represented as $c_2c_1$.
\begin{table}[h!]
%\begin{sidewaystable}[h!]
  \begin{center}
    \caption{Multiplication table for $251\times239=59989$ in binary.}
    \label{tab:table2}
    \resizebox{\textwidth}{!}{
    \begin{tabular}{ccccccccccccccccc}
      \hline
          & $2^{15}$ & $2^{14}$ & $2^{13}$ & $2^{12}$ & $2^{11}$ & $2^{10}$ & $2^{9}$ & $2^{8}$& $2^{7}$ & $2^{6}$ & $2^{5}$ & $2^{4}$ & $2^{3}$ & $2^{2}$ & $2^{1}$ & $2^{0}$ \\\hline\hline
    p     &          &          &          &          &          &          &         &          &  $1$ & $p_{6}$ & $p_{5}$ & $p_{4}$      & $p_{3}$ & $p_{2}$ & $p_{1}$ & $1$ \\
    q     &&&&&&&& &   $1$  &   $q_{6}$  &  $q_{5}$ & $q_{4}$ & $q_{3}$ & $q_{2}$ & $q_{1}$  & $1$\\\hline
     &&&&&&&& & 1 & \multicolumn{1}{:c}{$p_{6}$} & $p_{5}$ & $p_{4}$ & \multicolumn{1}{:c}{$p_{3}$} & $p_{2}$ & $p_{1}$ &\multicolumn{1}{:c}{$1$} \\
&&&&&&&& $q_1$ & $p_{6}q_{1}$ &  \multicolumn{1}{:c}{$p_{5}q_{1}$}   & $p_{4}q_{1}$ & $p_{3}q_{1}$ & \multicolumn{1}{:c}{$p_{2}q_{1}$} & $p_{1}q_{1}$ & $q_{1}$  &\multicolumn{1}{:c}{}  \\
&&&&&&&  \multicolumn{1}{:c}{$q_2$} & $p_{6}q_{2}$ &$p_{5}q_{2}$ & \multicolumn{1}{:c}{$p_{4}q_{2}$} & $p_{3}q_{2}$ & $p_{2}q_{2}$ & \multicolumn{1}{:c}{$p_{1}q_{2}$} & $q_{2}$      &          &  \multicolumn{1}{:c}{}  \\
&&&&&& $q_{3}$ & \multicolumn{1}{:c}{$p_{6}q_{3}$} & $p_{5}q_{3}$ & $p_{4}q_{3}$ & \multicolumn{1}{:c}{$p_{3}q_{3}$} & $p_{2}q_{3}$ & $p_{1}q_{3}$ & \multicolumn{1}{:c}{$q_3$} &              &          & \multicolumn{1}{:c}{}   \\
&&&&&$q_{4}$ & $p_{6}q_{4}$ &  \multicolumn{1}{:c}{$p_{5}q_{4}$} & $p_{4}q_{4}$ &$p_{3}q_{4}$ & \multicolumn{1}{:c}{$p_{2}q_{4}$} & $p_{1}q_{4}$ & $q_{4}$ & \multicolumn{1}{:c}{}&              &          &   \multicolumn{1}{:c}{} \\
&&&& \multicolumn{1}{:c}{$q_{5}$}&$p_{6}q_{5}$& $p_{5}q_{5}$ &  \multicolumn{1}{:c}{$p_{4}q_{5}$} & $p_{3}q_{5}$ & $p_{2}q_{5}$ & \multicolumn{1}{:c}{$p_{1}q_{5}$} & $q_{5}$ & &\multicolumn{1}{:c}{} &              &          & \multicolumn{1}{:c}{}\\
&&&$q_{6}$& \multicolumn{1}{:c}{$p_{6}q_{6}$}&$p_{5}q_{6}$&$p_{4}q_{6}$&  \multicolumn{1}{:c}{$p_{3}q_{6}$} & $p_{2}q_{6}$ & $p_{1}q_{6}$ & \multicolumn{1}{:c}{$q_{6}$} &  & & \multicolumn{1}{:c}{}  &              &          & \multicolumn{1}{:c}{}   \\
&&$1$&$p_{6}$& \multicolumn{1}{:c}{$p_{5}$}& $p_{4}$ & $p_{3}$ &  \multicolumn{1}{:c}{$p_{2}$} & $p_{1}$ &1 &\multicolumn{1}{:c}{} & & & \multicolumn{1}{:c}{} &              &          & \multicolumn{1}{:c}{}   \\
 &$c_{11}$&$c_{10}$&$c_{9}$& \multicolumn{1}{:c}{$c_{8}$}&$c_{7}$&$c_{6}$& \multicolumn{1}{:c}{$c_{5}$}&$c_{4}$&     $c_{3}$    & \multicolumn{1}{:c}{} & $c_{2}$ & $c_{1}$ &\multicolumn{1}{:c}{} & & & \multicolumn{1}{:c}{}   \\\hline
 &1&1&1&\multicolumn{1}{:c}{0}&1&0&\multicolumn{1}{:c}{1}& 0& 0 & \multicolumn{1}{:c}{1} & 0 & 1 & \multicolumn{1}{:c}{0} & 1 & 0 & \multicolumn{1}{:c}{1}   \\\hline
    \end{tabular}
 }   
  \end{center}
%\end{sidewaystable}  
\end{table}
\subsection{Factoring $N = 376289 = 659\times 571$}
The following table shows the carries $c_{14}c_{13}$, $c_{12}c_{11}c_{10}$, $c_{9}c_{8}c_7c_6$, $c_5c_4c_3$, $c_2c_1$ for corresponding blocks. There are overlaps in the column $2^{14}$.
\begin{table}[h!]
%\begin{sidewaystable}[h!]
  \begin{center}
    \caption{Multiplication table for $659\times 571=376289$ in binary.}
    \label{tab:table3}
\resizebox{\textwidth}{!}{\begin{tabular}{cccccccccccccccccccc}
      \hline
$2^{18}$ & $2^{17}$ & $2^{16}$ & $2^{15}$ & $2^{14}$ & $2^{13}$ & $2^{12}$ & $2^{11}$ & $2^{10}$ & $2^{9}$ & $2^{8}$& $2^{7}$ & $2^{6}$ & $2^{5}$ & $2^{4}$ & $2^{3}$ & $2^{2}$ & $2^{1}$ & $2^{0}$ \\\hline\hline
&&      &          &          &          &          &          &          &  $1$ &  $p_{8}$   & $p_{7}$& $p_{6}$ & $p_{5}$ & $p_{4}$      & $p_{3}$ & $p_{2}$ & $p_{1}$ & $1$ \\
&&       &&&&&&&$1$&$q_{8}$ &   $q_{7}$ &   $q_{6}$  &  $q_{5}$ & $q_{4}$ & $q_{3}$ & $q_{2}$ & $q_{1}$  & $1$\\\hline
&&   &&&&&&&$1$&$p_{8}$ & \multicolumn{1}{:c}{$p_{7}$} & $p_{6}$ & \multicolumn{1}{c:}{$p_{5}$} & $p_{4}$ & $p_{3}$ & $p_{2}$ & $p_{1}$ & \multicolumn{1}{:c}{$1$} \\
&&&&&&&& \multicolumn{1}{:c}{$q_1$}&$p_{8}q_{1}$& $p_{7}q_{1}$ & \multicolumn{1}{:c}{$p_{6}q_{1}$} &  $p_{5}q_{1}$   & $p_{4}q_{1}$ & \multicolumn{1}{:c}{$p_{3}q_{1}$} & $p_{2}q_{1}$ & $p_{1}q_{1}$ & $q_{1}$  &\multicolumn{1}{:c}{}  \\
&&&&&&& $q_2$&\multicolumn{1}{:c}{$p_{8}q_{2}$} & $p_{7}q_{2}$& $p_{6}q_{2}$ & \multicolumn{1}{:c}{$p_{5}q_{2}$} & $p_{4}q_{2}$ & $p_{3}q_{2}$ & \multicolumn{1}{:c}{$p_{2}q_{2}$} & $p_{1}q_{2}$ & $q_{2}$      &          &  \multicolumn{1}{:c}{}  \\
&&&&&& $q_{3}$ & $p_{8}q_{3}$& \multicolumn{1}{:c}{$p_{7}q_{3}$}& $p_{6}q_{3}$ & $p_{5}q_{3}$ & \multicolumn{1}{:c}{$p_{4}q_{3}$} & $p_{3}q_{3}$ & $p_{2}q_{3}$ & \multicolumn{1}{:c}{$p_{1}q_{3}$} & $q_3$ &              &          & \multicolumn{1}{:c}{}   \\
&&&&& \multicolumn{1}{:c}{$q_{4}$}& $p_{8}q_{4}$& $p_{7}q_{4}$ & \multicolumn{1}{:c}{$p_{6}q_{4}$} & $p_{5}q_{4}$ & $p_{4}q_{4}$ & \multicolumn{1}{:c}{$p_{3}q_{4}$} & $p_{2}q_{4}$ & $p_{1}q_{4}$ & \multicolumn{1}{:c}{$q_{4}$} & &              &          &  \multicolumn{1}{:c}{}  \\
&&&&$q_{5}$&\multicolumn{1}{:c}{$p_{8}q_{5}$}&$p_{7}q_{5}$&$p_{6}q_{5}$&\multicolumn{1}{:c}{ $p_{5}q_{5}$} & $p_{4}q_{5}$ & $p_{3}q_{5}$ & \multicolumn{1}{:c}{$p_{2}q_{5}$} & $p_{1}q_{5}$ & $q_{5}$ & \multicolumn{1}{:c}{} & & & &\multicolumn{1}{:c}{} \\
&&&$q_{6}$&$p_{8}q_{6}$&\multicolumn{1}{:c}{$p_{7}q_{6}$}&$p_{6}q_{6}$&$p_{5}q_{6}$& \multicolumn{1}{:c}{$p_{4}q_{6}$}& $p_{3}q_{6}$ & $p_{2}q_{6}$ &  \multicolumn{1}{:c}{$p_{1}q_{6}$} & $q_{6}$ &  & \multicolumn{1}{:c}{}& & & & \multicolumn{1}{:c}{} \\
&&\multicolumn{1}{:c}{$q_{7}$}&$p_{8}q_{7}$&$p_{7}q_{7}$&\multicolumn{1}{:c}{$p_{6}q_{7}$}&$p_{5}q_{7}$& \multicolumn{1}{c:}{$p_{4}q_{7}$}& $p_{3}q_{7}$ & $p_{2}q_{7}$ & $p_{1}q_{7}$ & \multicolumn{1}{:c}{$q_{7}$} &  & &  \multicolumn{1}{:c}{} & & && \multicolumn{1}{:c}{} \\
&$q_{8}$&\multicolumn{1}{:c}{$p_{8}q_{8}$}&$p_{7}q_{8}$&$p_{6}q_{8}$&$p_{5}q_{8}$& $p_{4}q_{8}$ & $p_{3}q_{8}$ &\multicolumn{1}{:c}{$p_{2}q_{8}$} & $p_{1}q_{8}$ & $q_{8}$ &\multicolumn{1}{:c}{}  & &  &  \multicolumn{1}{:c}{} & &   &&\multicolumn{1}{:c}{} \\
$1$&$p_{8}$&\multicolumn{1}{:c}{$p_{7}$}&$p_{6}$&$p_{5}$& \multicolumn{1}{:c}{$p_{4}$} & $p_{3}$ & $p_{2}$ & \multicolumn{1}{:c}{$p_{1}$} & 1 && \multicolumn{1}{:c}{} & & &\multicolumn{1}{:c}{} &  && & \multicolumn{1}{:c}{}\\
$c_{14}$&$c_{13}$&\multicolumn{1}{:c}{}&&$c_{9}$&\multicolumn{1}{:c}{$c_8$}&$c_7$&$c_{6}$&\multicolumn{1}{:c}{$c_{5}$}&$c_{4}$& $c_{3}$    & \multicolumn{1}{:c}{} & $c_{2}$ & $c_{1}$ &\multicolumn{1}{:c}{} & & &          & \multicolumn{1}{:c}{} \\
&&\multicolumn{1}{:c}{$c_{12}$}&$c_{11}$&$c_{10}$&\multicolumn{1}{:c}{}&&&\multicolumn{1}{:c}{}&&& \multicolumn{1}{:c}{} &&&\multicolumn{1}{:c}{} & & & &  \multicolumn{1}{:c}{}\\\hline
1&0&\multicolumn{1}{:c}{1}&1&0&\multicolumn{1}{:c}{1}&1&1&\multicolumn{1}{:c}{1}&0& 1& \multicolumn{1}{:c}{1}&1& 1 &\multicolumn{1}{:c}{0} & 0 & 0 & 0 & \multicolumn{1}{:c}{1}   \\\hline
    \end{tabular}
    }
  \end{center}
%\end{sidewaystable}  
\end{table}

\subsection{Range of Coefficients}
Define the lengths of $p$ and $q$ as $l_1$ and $l_2$, respectively. Let $l_1 = \frac{log(N)}{2} = O(log(N))$ and $l_2 = \frac{log(N)}{2} = O(log(N))$. Suppose each block contains 3 columns as in Table 2 for factoring 59989. Then the sum for one block is not larger than $(l_2+1)+2(l_2+1)+4(l_2+1)=7l_2+7$, assuming all unknown bits in $p$ and $q$ are 1's and all carries from the block on the right hand side are 1's. Thus, the length of the sum for the current block is at most $log(7l_2+7) = O(log(log(N)))$. Therefore, the length of the carry from the sum of current block is at most $log(7l_2+7)-3 = O(log(log(N)))$. Because the length of the carry plus the width of the block (which is 3 in this case) determines the range of the coefficients in the cost function, the maximum coefficient in the cost function corresponding to one block is $(2^{O(log(log(N)))})^2=O((log(N))^2)$. This square comes from transforming the equation for each block to a square that makes the equation hold. (See appendix A.2 for examples of these equations.) There are approximately $\frac{log(N)}{3}$ blocks in total for this example, such that the coefficient in the combined cost function containing all cost functions for each block is no larger than  $\frac{log(N)}{3} * O((log(N))^2) = O((log(N))^3)$. Note that for the majority of cases, this becomes $O((log(N))^2)$ because most of terms in different blocks are different. Since the variable replacement only effects the scale of the range of the coefficients linearly, the coefficients of the final quadratic cost function are polynomially large with regard to the size of $N$, the number to be factored.
\end{document}